\documentclass[webpdf,contemporary,large,namedate]{oup-authoring-template}%

\usepackage{algorithmicx}
\usepackage{lmodern}
\usepackage{setspace}

\begin{document}
\journaltitle{TBD}
\DOI{TBD}
\copyrightyear{2025}
\pubyear{2025}
\access{Advance Access Publication Date: Day Month Year}
\appnotes{Preprint}
\firstpage{1}

\title[Improving spliced alignment]{Improving spliced alignment by modeling splice sites with deep learning}
\author[1,2]{Siying Yang}
\author[1,2]{Neng Huang}
\author[1,2,3,$\ast$]{Heng Li\ORCID{0000-0003-4874-2874}}
\address[1]{Department of Biomedical Informatics, Harvard Medical School, 10 Shattuck St, Boston, MA 02215, USA}
\address[2]{Department of Data Science, Dana-Farber Cancer Institute, 450 Brookline Ave, Boston, MA 02215, USA}
\address[3]{Broad Insitute of MIT and Harvard, 415 Main St, Cambridge, MA 02142, USA}
\corresp[$\ast$]{Corresponding author. \href{mailto:hli@ds.dfci.harvard.edu}{hli@ds.dfci.harvard.edu}}


\abstract{
\sffamily\footnotesize
\textbf{Motivation:}
Spliced alignment refers to the alignment of messenger RNA (mRNA) or protein sequences to eukaryotic genomes.
It plays a critical role in gene annotation and the study of gene functions.
Accurate spliced alignment demands sophisticated modeling of splice sites,
but current aligners use simple models, which may affect their accuracy given dissimilar sequences.
\vspace{0.5em}\\
\textbf{Results:}
We implemented minisplice to learn splice signals with a one-dimensional convolutional neural network (1D-CNN)
and trained a model with 7,026 parameters for vertebrate and insect genomes.
It captures conserved splice signals across phyla and reveals GC-rich introns specific to mammals and birds.
We used this model to estimate the empirical splicing probability for every {\tt GT} and {\tt AG} in genomes,
and modified minimap2 and miniprot to leverage pre-computed splicing probability during alignment.
Evaluation on human long-read RNA-seq data and cross-species protein datasets showed
our method greatly improves the junction accuracy especially for noisy long RNA-seq reads
and proteins of distant homology.
\vspace{0.5em}\\
\textbf{Availability and implementation:}
\url{https://github.com/lh3/minisplice}
}

\maketitle

\section{Introduction}

RNA splicing is the process of removing introns from precursor mRNAs (pre-mRNAs).
It is widespread in eukaryotic genomes~\citep{Keren:2010aa}.
In human, for example, each protein-coding gene contains 9.4 introns on average;
$>$98\% of introns start with {\tt GT} on the genome (or more precisely {\tt GU} on the pre-mRNA)
and $>$99\% end with {\tt AG}~\citep{Sibley:2016vh}.
On the other hand, there are hundreds of millions of di-nucleotide {\tt GT} or {\tt AG}
in the human genome.
Only $\sim$0.1\% of them are real splice sites.
Identifying real splice sites, which is required for gene annotation,
is challenging due to the low signal-to-noise ratio.

To annotate splice sites and genes in a new genome,
we can perform RNA sequencing (RNA-seq) and align mRNA sequences to the target genome.
This approach does not work well for genes lowly expressed in sequenced tissues.
A complementary strategy is to align mRNA or protein sequences from other species to the target genome.
Spliced alignment through introns is essential in both cases.

\begin{figure}[b]
\centering
\includegraphics[width=.7\columnwidth]{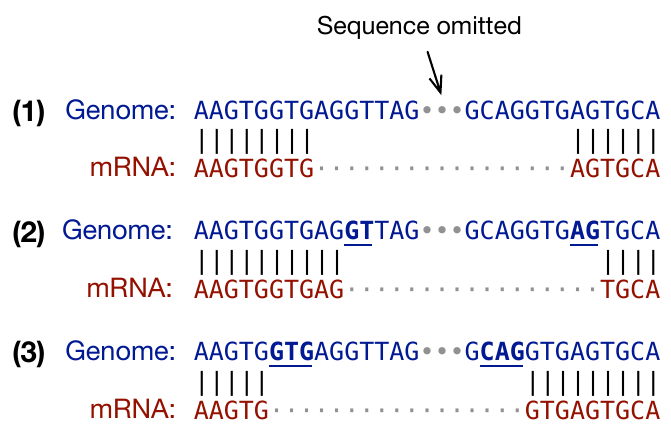}
\caption{Ambiguity in spliced alignment.
The same genome-mRNA sequence pair can be aligned differently without mismatches or gaps.}\label{fig:1}
\end{figure}

It is important to look for splice signals during spliced alignment
as the residue alignment around a splice site can be ambiguous.
For example, the three alignments in Fig.~\ref{fig:1} are equally good if we ignore splice signals.
However, as the putative intron in alignment (1) does not match the {\tt GT..AG}
signal, it is unlikely to be real.
While both (2) and (3) match the signal,
alignment (3) is more probable because it fits the splice consensus {\tt GTR..YAG} better~\citep{Irimia:2008aa,Iwata:2011aa},
where ``{\tt R}'' stands for an {\tt A} or a {\tt G} base and ``{\tt Y}'' for {\tt C} or {\tt T}.
In this toy example, the query sequence matches the reference perfectly in all three cases.
On real data, spliced aligners may introduce extra mismatches and gaps to reach splice sites.
The splice model has a major influence on the final alignment especially for diverged seqeences
when aligners need to choose between multiple similarly scored alignments around splice junctions.

Position weight matrix (PWM) is a classical method for modeling splice signals~\citep{Staden:1984aa}.
It however does not perform well because it cannot capture dependencies between positions~\citep{Burge:1997uu}
or model regulatory motifs such as branch points that do not have fixed positions.
Many models have been developed to overcome the limitation of PWM~\citep{Capitanchik:2025aa}.
In recent years, deep learning is gaining attraction
and has been shown to outperform traditional methods~\citep{Zhang:2018aa,DBLP:journals/access/DuYDZZL18,Albaradei:2020aa}.
Early deep learning models are small with only a few 1D-CNN layers.
Later models are larger, composed of residual blocks~\citep{Jaganathan:2019aa,Zeng:2022aa,Xu:2024aa,Chao:2024aa} or transformer blocks~\citep{You:2024aa,Chen:2024aa}.
It is also possible to fine tune genomic large-language models for splice site prediction~\citep{Nguyen:2023aa,Dalla-Torre:2025aa,Brixi2025.02.18.638918}.
Developed for general purposes, large-language models are computationally demanding and may be overkill if we just use them to predict splice sites.

At the same time, Helixer~\citep{Holst2023.02.06.527280} and Tiberius~\citep{Gabriel:2024aa}
combined Hidden Markov Models (HMM) and deep-learning models for \emph{ab initio} gene prediction and achieved high accuracy.
They however do not report alternative isoforms and are not suitable for genomes with frequent alternative splicing.
Mainstream gene annotation pipelines such as Ensembl and NCBI EGAP/EGAPx still heavily rely on alignment.

While qualified spliced aligners all look for the {\tt GT..AG} splice signal,
they model additional flanking sequences differently.
Intra-species mRNA-to-genome aligners such as BLAT~\citep{Kent:2002jk}, GMAP~\citep{Wu:2005vn} and Splign~\citep{Kapustin:2008tq} often do not model extra sequences beyond {\tt GT..AG}
because alignment itself provides strong evidence and ambiguity shown in Fig~\ref{fig:1} is rare.
Minimap2~\citep{Li:2018ab} prefers the {\tt GTR..YAG} consensus~\citep{Irimia:2008aa}.
This helps to improve the alignment of noisy long RNA-seq reads.
GSNAP~\citep{Wu:2010uq} integrates MaxEnt~\citep{Yeo:2004aa} for scoring novel splice sites.
Protein-to-genome aligners tend to employ better models due to more ambiguous alignment given distant homologs.
Miniprot~\citep{Li:2023ab} considers rarer {\tt GC..AG} and {\tt AT..AC} splice sites and optionally prioritizes on the {\tt G|GTR..YNYAG} consensus
common in vertebrate and insect~\citep{Iwata:2011aa}, where ``{\tt |}'' indicates splice boundaries.
Exonerate~\citep{Slater:2005aa}, Spaln~\citep{Gotoh:2008aa,Iwata:2012aa,Gotoh:2024aa} and the original GeneWise~\citep{Birney:2004uy} use PWM.
GeneSeqer~\citep{Usuka:2000vi} and GenomeThreader~\citep{DBLP:journals/infsof/GremmeBSK05} apply more advanced models~\citep{Brendel:1998aa,Brendel:2004aa}.
Deep learning models have been applied to refining splice sites as a post-processing step~\citep{Chao:2024aa,Xia:2023aa}
but have not been integrated into spliced aligners.

In this article, we introduce minisplice, a command-line tool implemented in C,
that learns splice signals and scores candidate splice sites with a small 1D-CNN model.
We have modified minimap2 and miniprot to take the splice scores as input for improved spliced alignment.
Importantly, we aim to develop a simple model that is more capable than PWM and is still easy to deploy;
we do not intend to compete with the best splice models which are orders of magnitude larger.

\begin{figure}[tb]
\includegraphics[width=.48\textwidth]{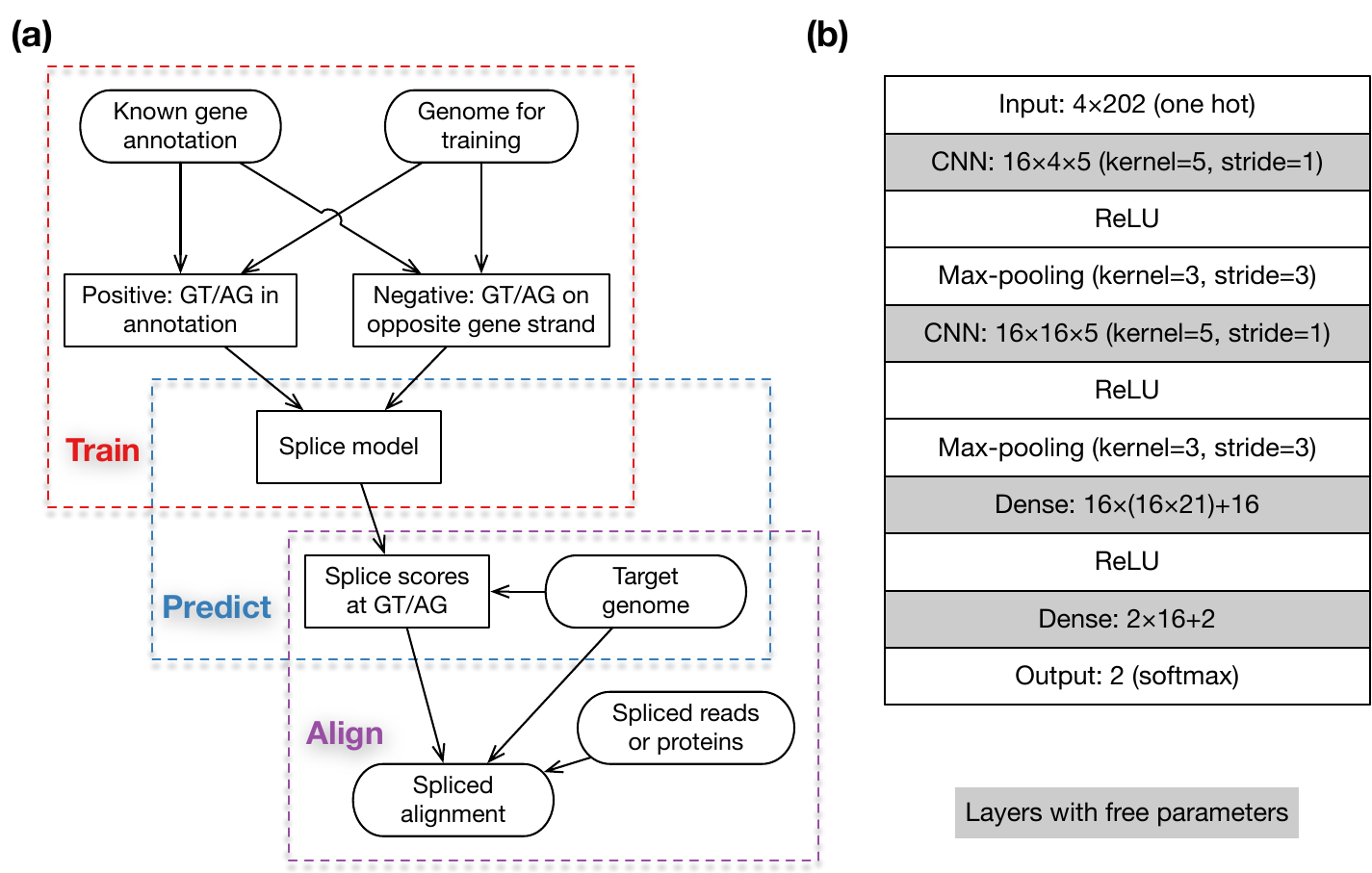}
\caption{Method overview. {\bf (a)} Overall workflow.
Training and prediction are done by minisplice.
Alignment is done by minimap2 or miniprot.
{\bf (b)} Default model architecture and parameterization, including tensor shapes.
For example, ``$4\times202$'' indicates the input is a $4\times202$ matrix;
``$16\times4\times5$'' suggests the 1D-CNN layer has a kernel size of 5 and outputs 16 features.
Shaded boxes contain free parameters.}\label{fig:wf}
\end{figure}

\section{Methods}

Our overall workflow consists of three steps: training, prediction and alignment (Fig.~\ref{fig:wf}a).
First, we train a deep learning model and transform scores outputted by the model
to empirical probabilities using known gene annotation.
Second, given a target genome to which mRNA or protein sequences will be aligned,
we predict the empirical probability of splice sites at each {\tt GT} or {\tt AG} in the genome
and output the logarithm-scaled splice scores to a file.
Third, when aligning mRNA sequences with minimap2~\citep{Li:2018ab} or aligning protein sequences with miniprot~\citep{Li:2023ab},
we feed the precomputed splice scores to the aligners which use the scores during dynamic-programming-based residue alignment.
This procedure will improve the accuracy around splice sites.
As we will show later, we can merge the training data from several vertebrates and insect
to obtain a model working well across phyla.
We do not need to train often.

\subsection{Generating training data}

Consistent with the literature, we call the 5'-end of an intron as the \emph{donor} site
and 3'-end as the \emph{acceptor} site.
Minisplice takes the genome sequence in the FASTA format
and gene annotation in the 12-column BED format (BED12) as input.
It provides a script to convert annotation in GTF or GFF3 format to BED12.
To generate training data, minisplice inspects each annotated donor site and labels it as a positive site if its sequence is {\tt GT};
similarly, minisplice labels a positive acceptor if it is an annotated {\tt AG}.
We ignore donor sites without {\tt GT} or acceptor sites without {\tt AG}
because other types of splice sites are rare~\citep{Sibley:2016vh}.
Miniprot and minimap2 still consider rarer {\tt GC..AG} or {\tt AT..AC} at the alignment step.

Due to potentially incomplete gene annotation in non-model organisms,
a {\tt GT} dinucleotide could be a real unannotated donor site.
To alleviate this issue, we label an unannotated {\tt GT} as a negative donor site
only if it comes from the opposite strand of an annotated gene~\citep{Chao:2024aa}.
In the rare case when two genes on opposite strands overlap with each other,
we ignore the overlapping region.
To balance positive and negative sites,
we randomly selsect a subset of negative sites such that the positive-to-negative ratio is 1:3.
For each positive or negative donor site,
we extract 100bp immediately before and after {\tt GT}.
The total length of sequences used for training is thus 202bp.
Negative acceptor sequences are generated similarly.
We also experimented 102bp and 302bp window sizes.

\subsection{Model architecture}

Minisplice uses a model with two 1D-CNN layers (Fig.~\ref{fig:wf}b).
The architecture is common among small models for splice site prediction~\citep{Zabardast:2023aa}.
The default model uses 16 features at both CNN layers and has 7,026 free parameters in total (sum of numbers in shaded boxes).
During development, we experimented alternative models
with different kernel sizes, more CNN or dense layers, optional dropout layers or more parameters.
We chose a relatively small model in the end as it is more efficient to deploy.
Because minimap2 and miniprot score each donor or acceptor site independently,
we also model splice sites independently.

\subsection{Training and testing}

We use 80\% of genes on the odd chromosomes or contigs for training and reserve the rest 20\% for validation.
We stop training if the validation cost increases over several epochs.
Recall that we intend to predict splice sites across the whole genome
but when generating training data, we downsample negative {\tt GT}/{\tt AG} to a small fraction.
To test the model accuracy in a setting closer to the prediction task,
we apply the trained model to every {\tt GT}/{\tt AG} on the even chromosomes or contigs
and compare the prediction to the known gene annotation to measure accuracy.
In comparison to training data, testing data may contain errors in known gene annotation:
missing junctions in the annotation would appear to be false positives (FPs),
while falsely annotated junctions would look false negatives (FNs).
It is not straightforward to compare accuracy across species.

\subsection{Transforming raw model scores to probabilities}

With the `softmax' operator at the end,
the model scores each candidate splice site with a number between 0 and 1, the higher the better.
This score is not a probability in particular when the property of the training data
is distinct from our intended application.
We need to transform this score to probability to work with the probability-based scoring system of miniprot.

We evenly divide raw model scores into $b$ bins (50 by default) such that
raw score $t\in[0,1)$ is assigned to bin $i=\lfloor tb\rfloor$, $i=0,1,\ldots,b-1$.
Let $P_i$ be the number of annotated splice sites scored to bin $i$
and $N_i$ be the number of unannotated {\tt GT}/{\tt AG} sites scored to bin $i$.
$P_i/(P_i+N_i)$ is the empirical probability of a candidate site in bin $i$ being real.
Let $P=\sum_i P_i$ and $N=\sum_i N_i$.
Given raw score $t$, the transformed score is
\begin{equation}\label{eq:s}
s(t)\triangleq 2\log_2\left(\frac{P_{\lfloor tb\rfloor}}{P_{\lfloor tb\rfloor}+N_{\lfloor tb\rfloor}}\cdot\frac{P+N}{P}\right)
\end{equation}
It computes the log odds of the probability estimated with the deep learning model
versus with the null model that assumes every {\tt GT}/{\tt AG} having equal probability of being real.
The $2\log_2$ scaling is imposed by BLOSUM scoring matrices~\citep{Henikoff:1992tk} which miniprot uses.

\subsection{Aligner integration}

Minimap2~\citep{Li:2018ab} uses the following equation for spliced alignment:
\begin{equation}\label{eq:splice}
\left\{\begin{array}{l}
H_{ij} = \max\{H_{i-1,j-1}+s(i,j),E_{ij},F_{ij},\tilde{E}_{ij}-a(i)\}\\
E_{i+1,j}= \max\{H_{ij}-q,E_{ij}\}-e\\
F_{i,j+1}= \max\{H_{ij}-q,F_{ij}\}-e\\
\tilde{E}_{i+1,j}= \max\{H_{ij}-d(i)-\tilde{q},\tilde{E}_{ij}\}\\
\end{array}\right.
\end{equation}
where $q$ is the gap open penalty, $e$ the gap extension penalty
and $s(i,j)$ gives the substitution score between the $i$-th position
on the reference and the $j$-th position on the query sequence.
$d(i)$ and $a(i)$ are the donor and acceptor scores, respectively, calculated with Eq.~(\ref{eq:s}).
Miniprot~\citep{Li:2023ab} uses a more complex equation which has the same donor and acceptor score functions.

\subsection{Implementation}

Minisplice is implemented in the C programming language with the only dependency
being zlib for reading compressed files.
It uses a deep-learning library we developed earlier for identifying human centromeric repeats~\citep{Li:2019aa}.
Minisplice outputs splice scores in a TAB-delimited format like:
\begin{verbatim}
chr2   4184146   +   A   9
chr2   4184167   +   A   -5
chr2   4184191   -   D   5
chr2   4184199   +   A   -5
chr2   4184213   +   D   3
\end{verbatim}
where the second column corresponds to the offset of the splice boundary
and the last column gives the splice score.
We modified minimap2 and miniprot to optionally take such a file as input
and use the splice scores during residue alignment.
Notably, minimap2 and miniprot do not directly depend on minisplice.
Users can provide splice scores estimated by other means in principle.

\section{Results}

We evaluated the accuracy of trained models using Receiver Operating Characteristic (ROC) curves
where we computed the true positive rate and false positive rate at different thresholds on raw model scores.
In the ROC plot, we focused on the region with sensitivity above 50\% and false positive rate below 10\%
because we intend to improve spliced alignment in this region.
For each curve, we calculated rAUC, which is the area the under the ROC curve restricted to and scaled by this region.
Due to scaling, ${\rm rAUC}\in[0,1]$.

To find a small model that is generalized to multiple species and is fast to deploy,
we experimented models under several settings.
Our final model is trained on six insect genomes from five orders and seven vertebrate genomes.
We did not train a plant model because we are less familiar with the plant phylogeny.

\begin{table}[!tb]
\caption{Datasets\label{tab:data}}
\begin{tabular*}{\columnwidth}{@{\extracolsep\fill}lll@{\extracolsep\fill}}
\toprule
Label & Species & Accession \\
\midrule
human\dag      & \emph{Homo sapiens}               & GCA\_000001405.29 \\
mouse\dag*     & \emph{Mus musculus}               & GCA\_000001635.9 \\
chicken\dag*   & \emph{Gallus gallus}              & GCA\_016699485.1 \\
zebrafish\dag* & \emph{Danio rerio}                & GCA\_000002035.4 \\
fruitfly\dag*  & \emph{Drosophila melanogaster}    & GCA\_000001215.4 \\
mosquito\dag   & \emph{Anopheles gambiae}          & GCA\_943734735.2 \\
mCanLup*       & \emph{Canis lupus baileyi}        & GCF\_048164855.1 \\
mLagAlb*       & \emph{Lagenorhynchus albirostris} & GCF\_949774975.1 \\
bAccGen*       & \emph{Astur gentilis}             & GCF\_929443795.1 \\
bAnaAcu        & \emph{Anas acuta}                 & GCF\_963932015.1 \\
bTaeGut        & \emph{Taeniopygia guttata}        & GCF\_048771995.1 \\
rEmyOrb        & \emph{Emys orbicularis}           & GCF\_028017835.1 \\
aDenEbr        & \emph{Dendropsophus ebraccatus}   & GCF\_027789765.1 \\
fCarCar*       & \emph{Carassius carassius}        & GCF\_963082965.1 \\
fPunPun        & \emph{Pungitius pungitius}        & GCF\_949316345.1 \\
sMobHyp        & \emph{Mobula hypostoma}           & GCF\_963921235.1 \\
icTenMoli*     & \emph{Tenebrio molitor}           & GCF\_963966145.1 \\
idCalVici*     & \emph{Calliphora vicina}          & GCF\_958450345.1 \\
idStoCalc      & \emph{Stomoxys calcitrans}        & GCF\_963082655.1 \\
ihPlaCitr*     & \emph{Planococcus citri}          & GCF\_950023065.1 \\
ilCydFagi*     & \emph{Cydia fagiglandana}         & GCF\_963556715.1 \\
ilOstNubi      & \emph{Ostrinia nubilalis}         & GCF\_963855985.1 \\
iyBomTerr*     & \emph{Bombus terrestris}          & GCF\_910591885.1 \\
iyVesCrab      & \emph{Vespa crabro}               & GCF\_910589235.1 \\
\botrule
\end{tabular*}
\begin{tablenotes}\setlength\itemsep{0.0em}
Ensembl or Gencode annotations were used for model organisms (marked by ``\dag'');
RefSeq annotations used for non-model organisms whose labels
follow the naming standard developed by the Darwin Tree of Life Project:
prefix ``m'' stands for mammals,
``b'' for birds,
``r'' for reptiles,
``a'' for amphibians,
``f'' for fish,
``s'' for sharks,
``ic'' for order \emph{Coleoptera} (beetles),
``id'' for \emph{Diptera} (flies),
``ih'' for \emph{Hemiptera} (true bugs),
``il'' for \emph{Lepidoptera} (butterflies and moths),
``iy'' for \emph{Hymenoptera} (bees and ants).
Species marked by ``*'' are used for training cross-species models.
\end{tablenotes}
\end{table}

\subsection{Datasets}

We acquired the genome sequences and gene annotations for six model organisms and 16 non-model organisms (Table~\ref{tab:data}).
For model organisms, only chicken is annotated with the Ensembl pipeline;
others are annotated by third parties.
For non-model organisms, we intentionally chose species that have PacBio HiFi assemblies
and are annotated by both RefSeq and Ensembl if possible.
Only aDenEbr and sMobHyp do not have Ensembl annotations.

\subsection{Intra-species training}

\begin{figure}[tb]
\includegraphics[width=\columnwidth]{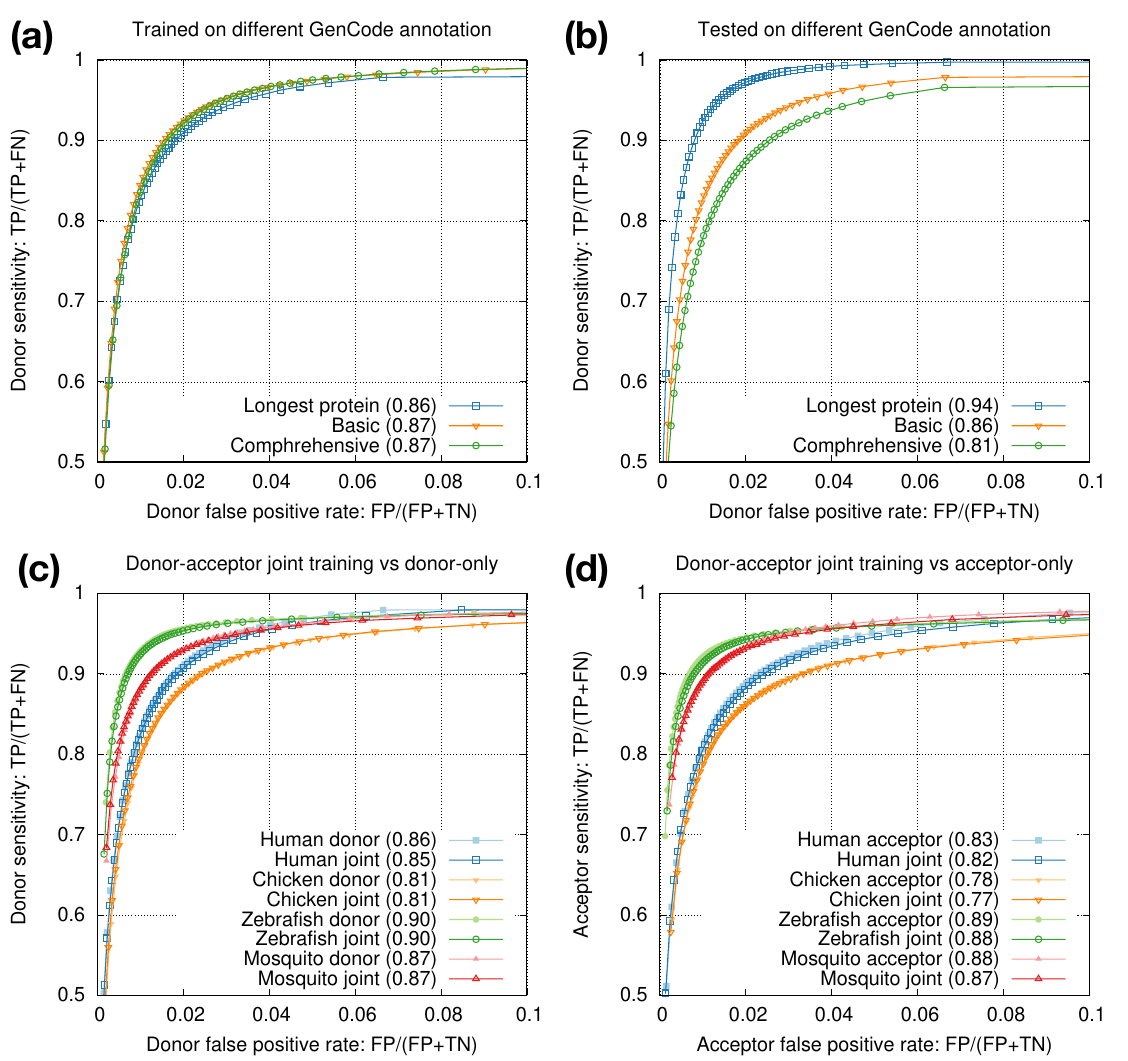}
\caption{Intra-species training.
{\bf (a)} Models trained on different human annotations
and tested on the Gencode basic annotation.
Numbers in parentheses denote rAUC, which is the Area Under the Curve restricted to the plotted region.
{\bf (b)} Model trained from the longest human protein-coding transcripts
and tested on different human annotations.
{\bf (c)} Donor-only training versus joint donor-acceptor training.
{\bf (d)} Acceptor-only training versus joint donor-acceptor training.}\label{fig:3}
\end{figure}

Gencode provides a smaller set of ``basic'' gene annotations and a larger set of ``comprehensive'' annotations.
We also have an option to select the longest transcript of each protein-coding gene for high-fidelity splice sites.
We found training data had minor effect on testing accuracy (Fig.~\ref{fig:3}a)
but annotations used for testing had larger effect (Fig.~\ref{fig:3}b).
We decided to train on splice sites from the longest protein-coding transcripts for training as they are most accurate,
and to test on basic annotations because non-model organisms probably do not have annotations comparable to comprehensive Gencode annotations.

The two central bases in donor training data are always {\tt GT}
and the two bases in acceptor training data are always {\tt AG}.
We speculated 1D-CNN models could easily learn the difference, so we mixed donor and acceptor training data
and trained one joint model for each species.
The joint models achieved nearly the same accuracy as separate donor or acceptor models (Fig.~\ref{fig:3}c,d).
In later experiments, we thus always trained one joint model to simplify the training process.

\subsection{Cross-species training}

\begin{figure}[bt]
\includegraphics[width=\columnwidth]{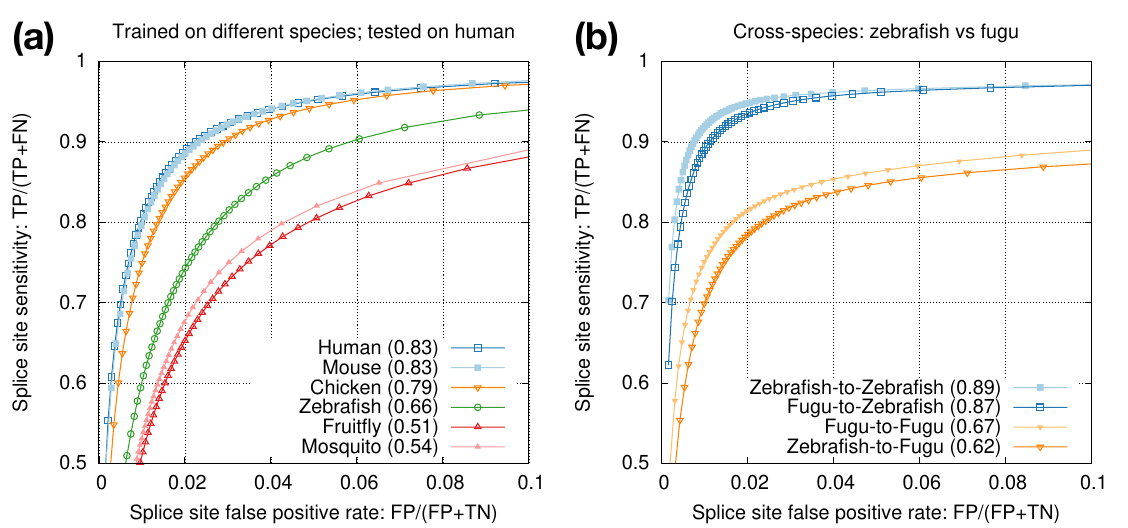}
\caption{Cross-species training.
{\bf (a)} Models trained from different species and tested on human.
{\bf (b)} Intra- versus cross-species training with mosquito and fruitfly.}\label{fig:4}
\end{figure}

Our end goal is to improve spliced alignment accuracy for species without known annotations.
Training and prediction are often applied to different species.
To test how well a model trained from one species can predict splice sites in a different species,
we applied models trained from model organisms to human (Fig.~\ref{fig:4}a).
We can see the test accuracy drops quickly with increased evolution distance.
The mouse model is almost good as the human model because mammals are closely related.
We also applied the mosquito model to the fruitfly genome (Fig.~\ref{fig:4}b).
The test accuracy is lower than the accuracy we obtain with the fruitfly model,
but the drop is much smaller in comparison to applying the mosquito model to human.
These experiments suggested we can achieve reasonable accuracy with a model trained from a closely related species.

\begin{figure}[tb]
\includegraphics[width=\columnwidth]{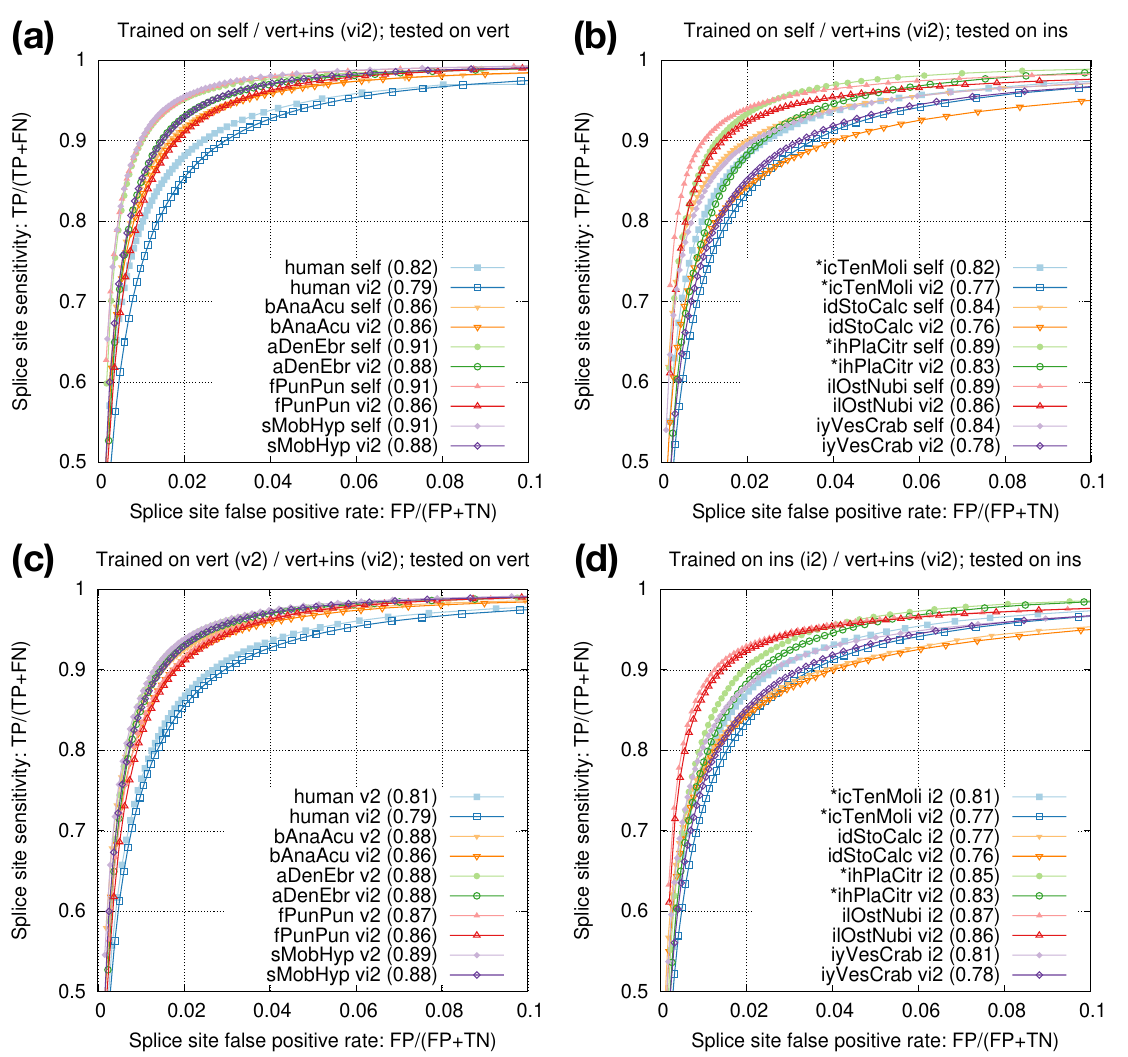}
\caption{Training from multiple species.
{\bf (a)} Accuracy of model vi2 on vertebrate genomes. vi2 is trained from six vertebrate and seven insect genomes.
{\bf (b)} Accuracy of vi2 on insect genomes. Odd chromosomes in starred species are used for training.
Testing is applied to even chromosomes only.
{\bf (c)} Comparison between vi2 and a model trained from vertebrate genomes (v2).
{\bf (d)} Comparison between vi2 and a model trained from insect genomes (i2).}\label{fig:5}
\end{figure}

\begin{figure}[bt]
\includegraphics[width=\columnwidth]{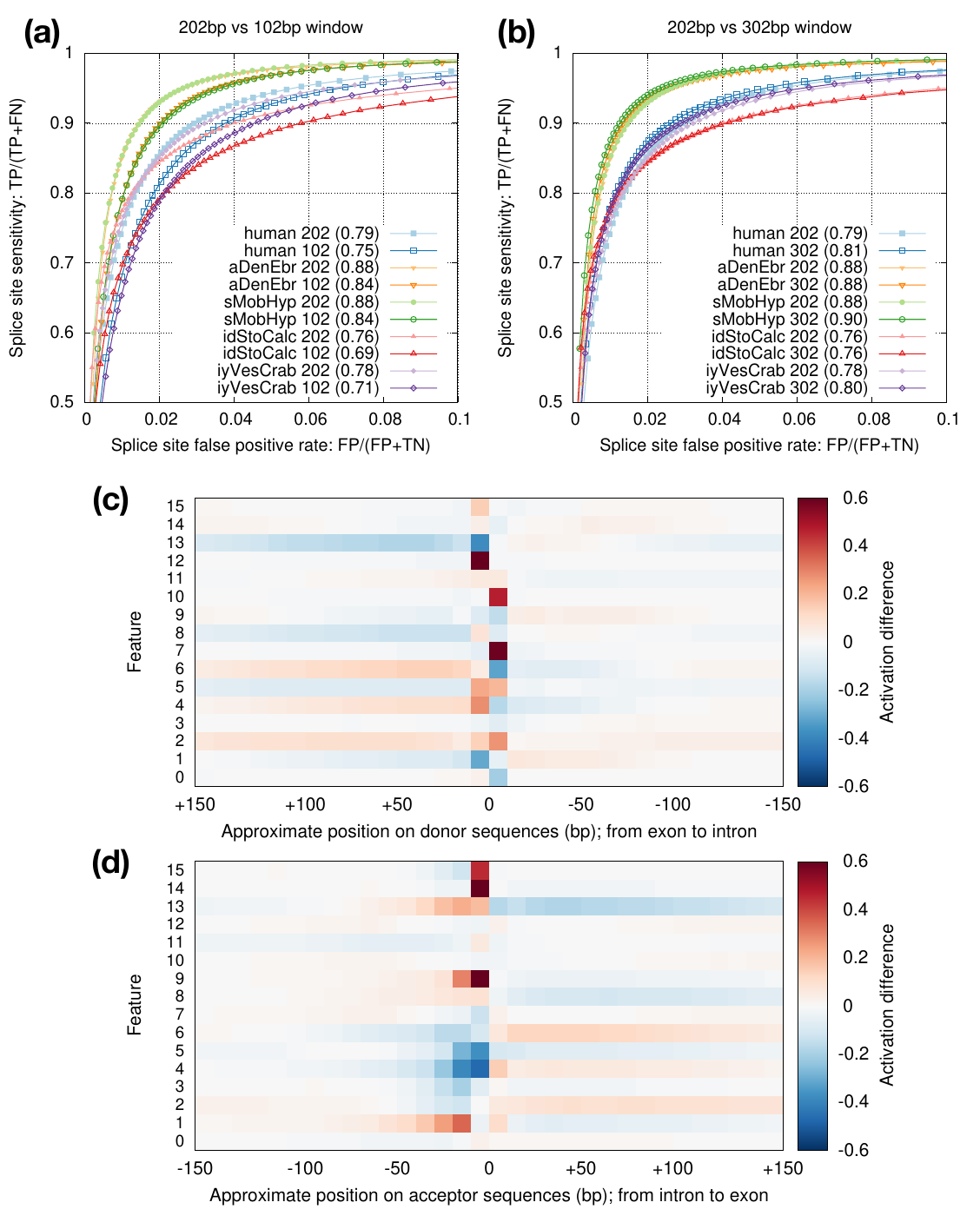}
\caption{Effect of window size.
{\bf (a)} Comparison between the default 202bp window size ($\pm$100bp around {\tt GT}/{\tt AG}) and 102bp.
{\bf (b)} Comparison between 202bp and 302bp window sizes.
{\bf (c)} Activation differences between positive donor sites and negative {\tt GT} sites at the last max-pooling layer.
Positive coordinates correspond to exonic bases.
With 302bp window, this layer consists of $16\times32$ cells which are non-negative due to the ReLU function.
Given a set of sequences, the activation rate of a cell is the frequency of the cell being positive (i.e. non-zero).
The heat shows the rate difference between positive and negative sets.
Dark red indicates more frequent activation among positive donor sites; dark blue is the opposite.
{\bf (d)} Activation difference for acceptor sites.}\label{fig:6}
\end{figure}

We went a step further by combining the training data across multiple vertebrate and insect species
and derived model vi2, which was trained from six vertebrate and seven insect genomes marked in Table~\ref{tab:data}.
Although this model is not as accurate as the species-specific model trained from individual species itself (Fig.~\ref{fig:5}a,b),
it is better than applying an insect model to human (Fig.~\ref{fig:4}a).
Note that vi2 is not trained on amphibian or shark genomes but it still accurately predicts splice sites in aDenEbr and sMobHyp.
It is capturing common signals across large evolutionary distance while reducing overfitting to individual species.
We also trained a vertebrate-only model (v2) and an insect-only model (i2).
They outperformed vi2 for vertebrate and insect genomes, respectively (Fig.~\ref{fig:5}c,d),
but we deemed the improvement is small and outweighed by the convenience
of having one model across vertebrate and insect genomes.

\begin{figure}[bt]
\includegraphics[width=\columnwidth]{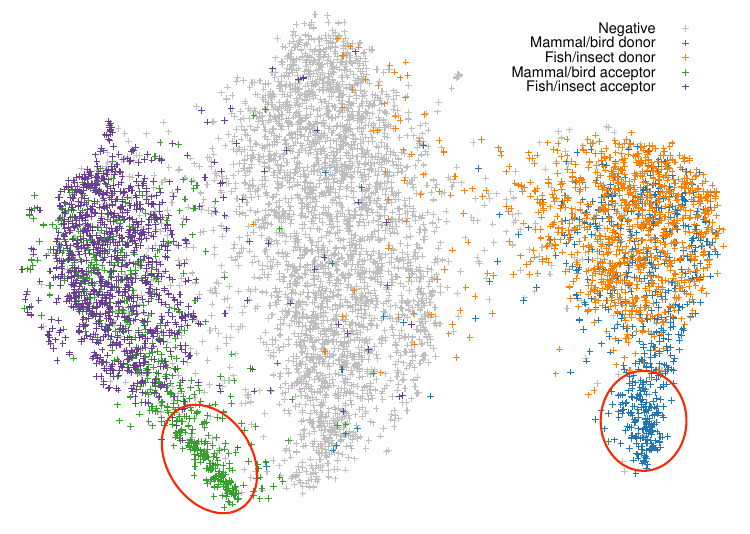}
\caption{UMAP of the last max-pooling layer.
Training data was downsampled to 20k splice sites with 10k positive and 10k negative sites.
}\label{fig:umap}
\end{figure}

\begin{figure}[bt]
\includegraphics[width=\columnwidth]{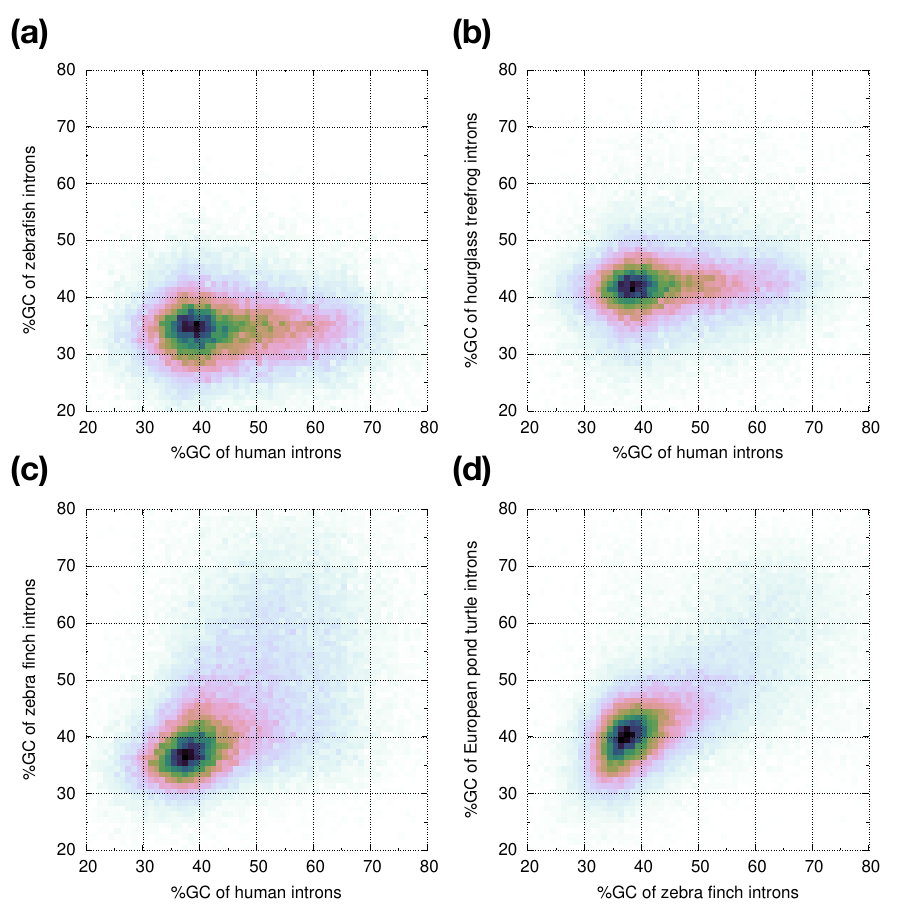}
\caption{GC content of homologous introns.
Vertebrate protein sequences from Swiss-Prot are aligned to each genome with miniprot.
Given two genomes, if a protein sequence is aligned with introns at the same position on the protein, the two aligned introns are considered homologous to each other.
{\bf (a)} Comparing percent GC content between homologous human and zebrafish introns.
{\bf (b)} Human versus hourglass tree frog (aDenEbr).
{\bf (c)} Human versus zebra finch (bTaeGut).
{\bf (d)} Zebra finch versus European pond turtle (rEmyOrb).
}\label{fig:gc}
\end{figure}

Model vi2 considers 202bp sequences around splice sites.
We tried to reduce the window size to 102bp but the accuracy dropped (Fig.~\ref{fig:6}a).
Increasing the window size to 302bp, on the other hand, only had a small effect (Fig.~\ref{fig:6}b).
To understand what signals the model is learning, we visualized the activation rates at the last max-pooling layer which has 16 features (Fig.~\ref{fig:6}c,d).
While some features such as 4 and 7 focus on signals at splice sites,
other features such as 2 and 13 likely capture compositions of introns and exons.
Intronic sequences around acceptors appear to provide more signals than around donors.
This might be related to branch points which are located within tens of basepairs upstream to acceptors.
The activation difference fades away beyond 100bp around splice sites especially within introns.
It is possibly why increasing window size from 202bp to 302bp has minor effect (Fig.~\ref{fig:6}b).

We further used the values at the last max-pooling layer to generate the UMAP of a random subset of training samples (Fig.~\ref{fig:umap}).
The UMAP separated the training samples into three large clusters which correspond to
positive acceptor sites, negative sites and positive donor sites, respectively.
Notably, some areas in the UMAP (the two red circles) only contain splice sites from mammals and birds but rarely from fish and insect.
We extracted splice sites in these areas
and found most of them came from GC-rich introns.
The presence of these areas might explain
why fish and insect models do not work well for mammals (Fig.~\ref{fig:4}a)
but mammalian models tend to work better for non-mammals~\citep{McCue:2024aa}.

To more directly investigate these GC-rich introns,
we extracted homologous intron pairs and compared their GC content (Fig.~\ref{fig:gc}).
We saw introns reaching 60\% GC in human but their homologs in zebrafish are $<$40\% in GC (Fig.~\ref{fig:gc}a).
Consistent with our earlier observation in UMAP, few zebrafish introns are GC-rich.
Although the GC content is overall higher in amphibians, not many introns can reach 60\% GC (Fig.~\ref{fig:gc}b).
Birds have high-GC introns like human (Fig.~\ref{fig:gc}c).
This is unrelated to warm- versus cold-blooded organisms per isochore theory~\citep{Bernardi:1985aa}
because reptiles also have GC-rich introns (Fig.~\ref{fig:gc}d).


GC-rich introns are known to exist in human~\citep{Wang:2011aa}
and may help stabilize RNA secondary structure~\citep{Zhang:2011aa}.
They are more likely to be retained than GC-poor introns~\citep{Sibley:2016vh},
and they tend to be located in the nuclear center~\citep{Choquet:2025aa}
which may be partly correlated with the higher CpG dentity in the center~\citep{Xie:2017aa,Tan:2018aa}.

\subsection{Improving spliced alignment}

\begin{table*}[!tb]
\caption{Effect of splice site models on protein-to-genome alignment\label{tab:eval}}
\begin{tabular*}{\textwidth}{@{\extracolsep\fill}llllllclllll@{\extracolsep\fill}}
\toprule
& \multicolumn{5}{c}{zebrafish proteins to human genome} && \multicolumn{5}{c}{mosquito proteins to fruitfly genome} \\[0.4em]
\cline{2-6}\cline{8-12}\\[-0.9em]
Aligner           & miniprot  & miniprot  & miniprot  & miniprot   & Spaln3    && miniprot & miniprot & miniprot & miniprot   & Spaln3 \\
Splice model      & GT-AG     & default   & extended  & minisplice & Tetrapod  && GT-AG    & default  & extended & minisplice & InsectDm \\
\midrule
\# prediced junc  & 168,030   & 164,094   & 160,935   & 164,860    & 144,544   && 30,279   & 28,780   & 27,307   & 28,722     & 24,538 \\
\# annotated junc & 144,495   & 146,898   & 147,880   & 157,654    & 132,394   && 24,022   & 24,465   & 24,203   & 27,107     & 21,161 \\
\% unannotated    & 14.01     & 10.48     & 8.11      & 4.37       & 8.41      && 20.68    & 14.99    & 11.37    & 5.62       & 13.76 \\
\% Base Sn        & 60.09     & 60.12     & 60.04     & 60.85      & 51.71     && 57.00    & 56.92    & 56.76    & 57.28      & 44.53 \\
\% Base Sp        & 94.33     & 95.49     & 96.09     & 97.30      & 94.37     && 98.39    & 98.65    & 98.64    & 99.28      & 97.37 \\
\botrule
\end{tabular*}
\begin{tablenotes}\setlength\itemsep{0.0em}
For the splice models, ``GT-AG'' only considers {\tt GT..AG},
``default'' considers {\tt GTR..YAG},
``extended'' additionally considers {\tt G|GTR..YYYAG},
and ``minisplice'' uses the ``vi2-7k'' model on top of ``extended''.
Miniprot was invoked with ``{\tt -I}'' which sets the maximum intron length to $\min\{\max\{3.6\sqrt{G},10^4\},3\times10^5\}$, where $G$ is the genome size.
Spaln3 was invoked with ``{\tt -LS -yS -yB -yZ -yX2}'' and with its own splice models.
${\rm\%unannotated}=1-{\rm\#annotated}/{\rm\#predicted}$.
\end{tablenotes}
\end{table*}

To evaluate the effect of splice models,
we aligned zebrafish proteins to the human genome with miniprot (Table~\ref{tab:eval})
and checked if an aligned junction is annotated in human.
Unannotated junctions are likely errors in the protein-to-genome alignment.
If we ask miniprot to only look for the minimal {\tt GT..AG} signal, 14.01\% of aligned junctions are not annotated in human.
This percentage drops with improved splice models
and reaches 4.37\% when minisplice scores are considered during alignment.
The similar trend is also observed when aligning mosquito proteins to the fruitfly genome.
This confirms the critical role of splice models in protein-to-genome alignment.

\begin{figure}[bt]
\includegraphics[width=\columnwidth]{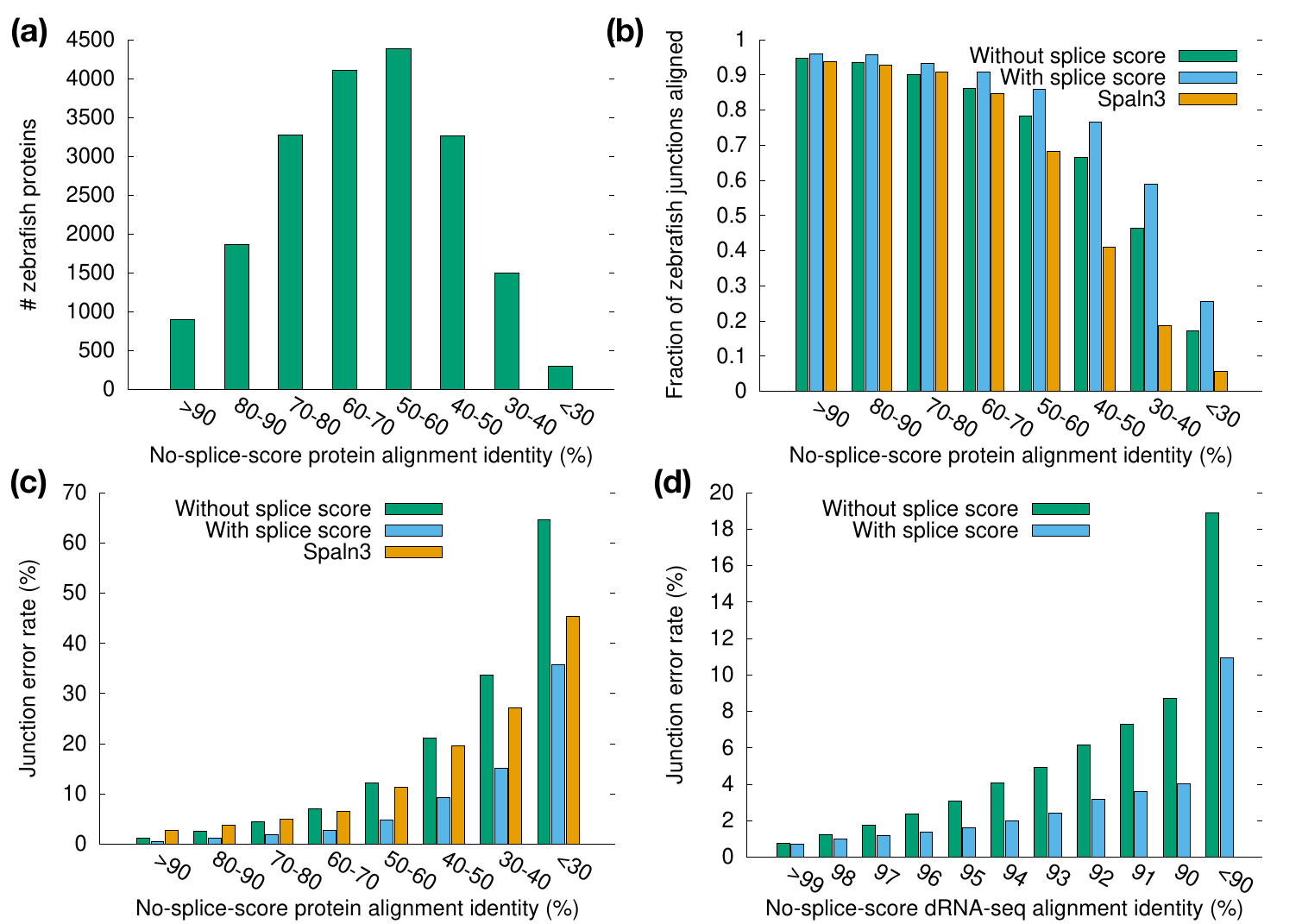}
\caption{Junction accuracy stratified by alignment identity.
{\bf (a)} Number of zebrafish proteins in each identity bin.
Identity is estimated from alignment of zebrafish proteins against the human genome under the default miniprot splice model.
{\bf (b)} Number of aligned junctions divided by the number of coding junctions in the zebrafish annotation.
{\bf (c)} Junction error rate of protein-to-genome alignment.
{\bf (d)} Junction error rate of RNA-seq alignment.
HG002 direct-RNA reads are aligned to the human genome with minimap2.}\label{fig:map}
\end{figure}

We stratified the zebrafish proteins by their alignment identity (Fig.~\ref{fig:map}a)
and inspected the alignment accuracy in each identity bin.
The fraction of aligned junctions and the junction accuracy both drop with alignment identity (Fig.~\ref{fig:map}b,c).
The minisplice model achieves higher accuracy across all bins, often halving the error rate in comparison to the default miniprot model.

We were evaluating junction accuracy.
The exon accuracy was 9--11\% lower for the zebrafish-to-human alignment.
We manually inspected some exon-specific errors and classified them into three cases.
First, an alignment did not reach the start of the protein sequence.
The first junction in the alignment could still be correct but the beginning of the first exon was often an error.
Second, 10.7\% of longest zebrafish proteins did not start with methionine ``M'', the start amino acid.
Third, zebrafish proteins starting with ``M'' may occasionally be mutated in human.
This is an interesting but rare case.
Overall, we intend to improve junction accuracy in this work but not the three cases above.
Junction accuracy serves our goal better.

The minisplice model also greatly reduces the junction error rate for direct RNA-seq reads (Fig.~\ref{fig:map}d),
though the reduction is less pronounced when the alignment identity reaches $>$98\%.
For this RNA-seq run with the latest the Nanopore RNA004 kit~\citep{Zheng2024.11.17.624050}, 74.6\% of reads are mapped with identity $>$98\%.
The overall junction error rate is marginally reduced from 1.4\% to 1.0\%.
The advantage of advanced splice models will be more noticeable for old data of lower quality, regions of high diversity or cross-species cDNA-to-genome alignment.

As to performance, minisplice precomputed splice scores for the human genome in 1.4 hours over 16 CPU threads.
This step only needs to be done once per genome.
Minimap2 and miniprot took $\sim$13 seconds upfront to load precomputed splice scores in the gzip format.
The following alignment speed was barely affected as looking up splice scores is much faster than alignment.

\section{Discussions}

Minisplice is a command-line tool to learn splice signals with 1D-CNN
and to predict the probability of splicing in the whole genome.
While neural networks have long been applied to the modeling of splice sites~\citep{Reese:1997aa},
minisplice represents the first effort to directly integrate deep learning into alignment algorithms.
The integration substantially improves the splice junction accuracy of minimap2 and miniprot,
and is likely to help downstream analyses such as transcriptome reconstruction and gene annotation.

A marked limitation of minisplice is that it only models {\tt GT..AG} splice sites.
Although miniprot considers {\tt GC..AG} and {\tt AT..AC} signals and minimap2 has recently adopted the same model,
they may still prefer {\tt GT..AG} well scored by minisplice and misalign introns with non-{\tt GT..AG} splicing.
In addition, minisplice predicts splice scores from the reference genome.
It is not aware of mutations in other samples that may alter splice signals.
This bias towards the reference genome may lead to misalignment in rare cases
but it is minor in comparison to annotation-guided alignment implemented in minimap2,
which is biased towards both the reference genome and the reference annotation.

If we could predict splice sites to 100\% accuracy, we would be close to solving the \emph{ab initio} gene finding problem.
However, the conserved splice signals only come from several bases around splice sites.
Given hundreds of millions of {\tt GT}/{\tt AG} in a mammalian genome, we would make many wrong predictions just based on the several bases.
The activation pattern at an internal layer suggests our model
draws power from long-range composition of exonic sequences in addition to short-range signals around splice sites.
The model may tend to classify a site as donor if the sequence on the left is similar to exonic sequences in composition.
Nonetheless, even at 1\% false positive rate, our model would predict 3.75 million ($=3\times10^9\times2/16\times1\%$) false donor sites on average,
an order of magnitude more than real donor sites.
We must consider additional information for more accurate prediction.
With $\ge$10kb windows and orders of magnitude more parameters, recent deep learning models
such as SpliceAI, Pangolin and DeltaSplice will learn composition better.
They may additionally see the promoter regions of many genes and species- or even tissue-specific regulatory elements.
It is not clear how much their signals come from splice sites.
At this point, methods combining HMM and deep learning~\citep{Gabriel:2024aa,Holst2023.02.06.527280}
may be more advantageous for coding sequences as they explicitly model gene structures and keep protein sequences in phase.

\section*{Acknowledgments}

We thank Kuan-Hao Chao for pointing us to the Splam and OpenSpliceAI models prior to their publication.

\section*{Author contributions}

H.L. conceived the project.
S.Y., N.H. and H.L. implemented the algorithms and analyzed the data.
S.Y. and H.L. drafted the manuscript.

\section*{Conflict of interest}

None declared.

\section*{Funding}

This work is supported by National Institute of Health grant R01HG010040 (to H.L.).

\section*{Data availability}

The minisplice source code is available at \url{https://github.com/lh3/minisplice}.
Pretrained models can be obtained from \url{https://doi.org/10.5281/zenodo.15446314}.

\bibliographystyle{apalike}
{\sffamily\small
\bibliography{minisplice}}

\end{document}